\documentclass[12pt]{article}
\usepackage{pic03}
\usepackage{hyperref}
\usepackage{url}
\usepackage{graphicx}

\begin{document}

\title{\bf STUDY OF QUARTIC BOSON VERTICES AT PHOTON COLLIDERS IN THE STANDARD MODEL AND BEYOND.}
\author{
I. Marfin, V. Mossolov, T. Shishkina  \\
{\em NCPHEP,  153 Bogdanovitcha str.,220040 Minsk}}
\maketitle

\baselineskip=14.5pt
\begin{abstract}
The production of several vector bosons in $\gamma\gamma$-interactions is the best place to
search directly for any anomalous behavior of triple and quartic
couplings. A few anomalous quartic boson operators are considered in the paper. 
A high precision of anomalous effects can be achieved by calculation of 
the lowest  order electroweak  corrections:
one-loop corrections, real photon and $Z$-boson emission.
\end{abstract}

\baselineskip=17pt

$ $

The multiple vector-boson production would be a
crucial test of the gauge structure of the Standard Model since the
triple and quartic vector-boson couplings  in this kind of
reaction are constrained by the $SU(2)\bigotimes U(1)$ gauge
invariance.  Any small deviation from the Standard Model predictions for these
couplings gives anomalous growth of the cross section with energy.
The quartic gauge boson couplings lead to direct  electroweak symmetry breaking,
in  particular in  the scalar  sector of  the theory  or, more
generally, to "new physics" of electroweak gauge bosons.
Since the processes $\gamma\gamma\rightarrow W^+W^-$ and
$\gamma\gamma\rightarrow W^+W^-Z$  at high energies
will give the unique possibility of quartic couplings investigation due to
relatively large cross sections and low background for
$WW$ and $WWZ$  productions \cite{c4}. 
Evaluating of anomalous contributions to cross sections of $WW$ and $WWZ$ productions requires
a calculation of  the lowest order electroweak  corrections.


Constructing the structures contained anomalous
quartic gauge boson couplings where at least one photon is involved 
one has to consider the operators with the the
lowest dimension of $6$ \cite{c2}. 
\begin{eqnarray}\label{a1}
\begin{array}{c}
\displaystyle {\cal L}_0 = -\frac{e^2}{16\Lambda^2}a_0F^{\mu\nu}
F_{\mu\nu}\bar{W}^{\alpha}\bar{W}_{\alpha}, ,\,\,\,\,\,
\displaystyle {\cal L}_c = -\frac{e^2}{16\Lambda^2}a_cF^{\mu\alpha}
F_{\mu\beta}\bar{W}^{\beta}\bar{W}^{\alpha}, \\  \\
\displaystyle \tilde{\cal L}_0 = -\frac{e^2}{16\Lambda^2}\tilde{a}_0
F^{\mu\alpha}\tilde{F}_{\mu\beta}\bar{W}^{\beta}\bar{W}^{\alpha}, 
\end{array}
\end{eqnarray}
where 
\begin{eqnarray}\label{a2}
\begin{array}{c}
\displaystyle \bar{W}_{\mu} = \left(\frac{1}{\sqrt{2}}(W^+_{\mu}+
W^-_{\mu}),\frac{i}{\sqrt{2}}(W^+_{\mu}-W^-_{\mu}),\frac{1}
{\cos{\theta_W}}Z_{\mu}\right), \\ \\
\displaystyle F_{\mu\nu} = \partial_{\mu}A_{\nu}
-   \partial_{\nu}A_{\mu}, \,\,\, \displaystyle W_{\mu\nu}^i = \partial_{\mu}W^i_{\nu}
-   \partial_{\nu}W^i_{\mu}, \,\,\,
\displaystyle \tilde{F}_{\mu\nu} = \frac{1}{2}
\epsilon_{\mu\nu\rho\sigma}F^{\rho\sigma}.
\end{array}
\end{eqnarray}
The scale $\Lambda$ ($\sim 80$ GeV) is introduced to keep the coupling constant
$a_i$ dimensionless \cite{c3}.  As one can see the operators ${\cal L}_0$ and ${\cal L}_c$
are $C$-, $P$-, $CP$-invariant. ${\cal L}_n$
violates both $C$- and  $CP$-invariance.   $\tilde{{\cal L}}_0$ is the $P$- and $CP$-violating operator.
$\tilde{{\cal L}}_n$ conserves $CP$-invariance but violates $C$- and $P$-invariance separately.

\begin{figure}[h!]
  \centerline{\hbox{ \hspace{0.1cm}
    \includegraphics[width=8.5cm]{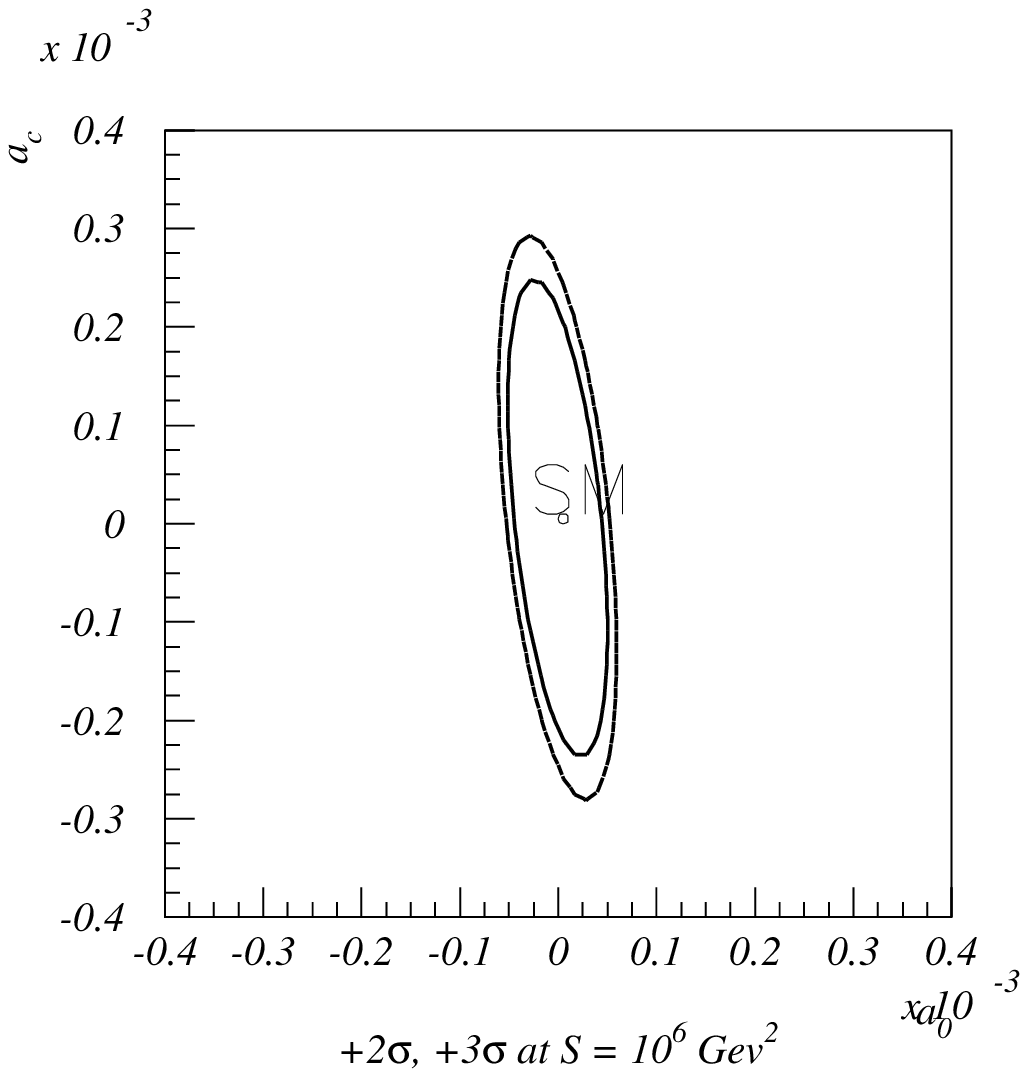}
    \hspace{0.1cm}
    \includegraphics[width=8.5cm]{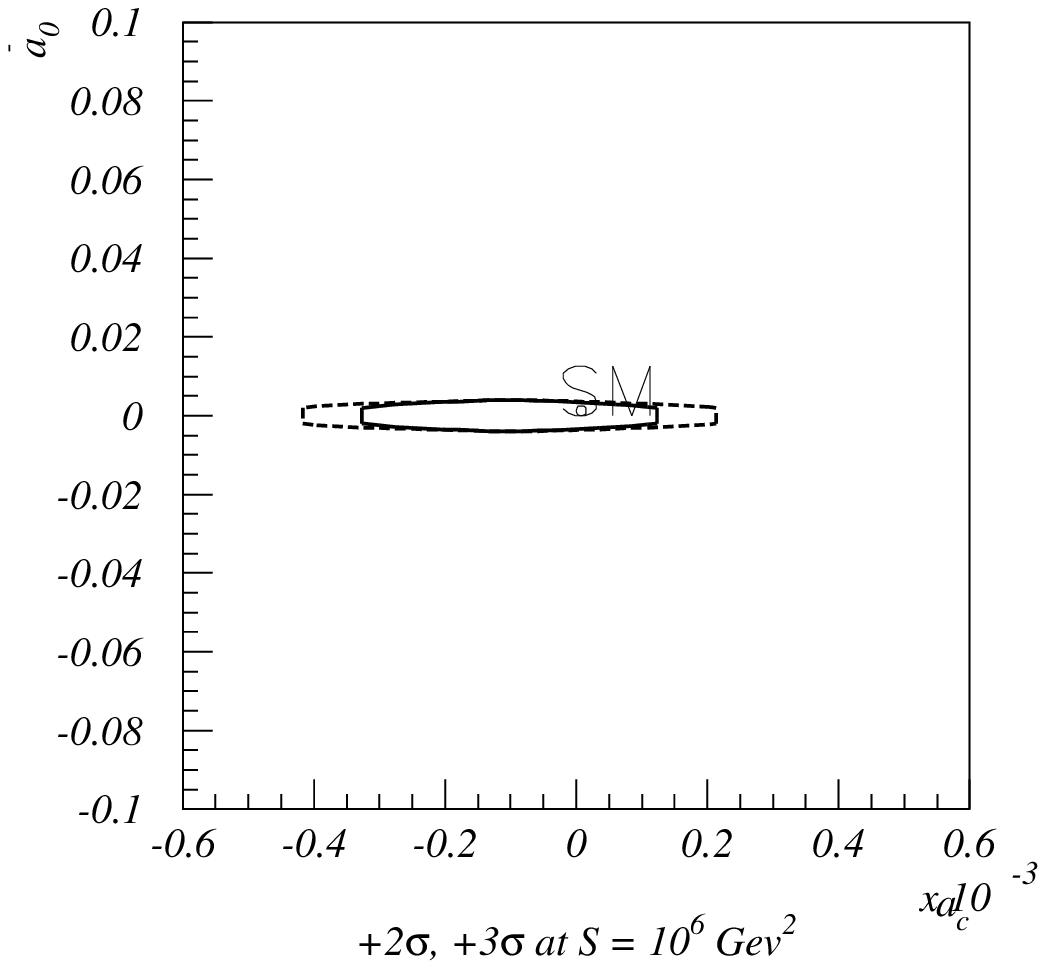}
    }
  }
 \caption{\it
      Contour plots for +2$\sigma$ and +3$\sigma$ deviations of $\sigma(\gamma\gamma\rightarrow W^+W^-)$
    \label{p1} }
\end{figure}

\begin{figure}[h!]
  \centerline{\hbox{ \hspace{0.1cm}
    \includegraphics[width=8.5cm]{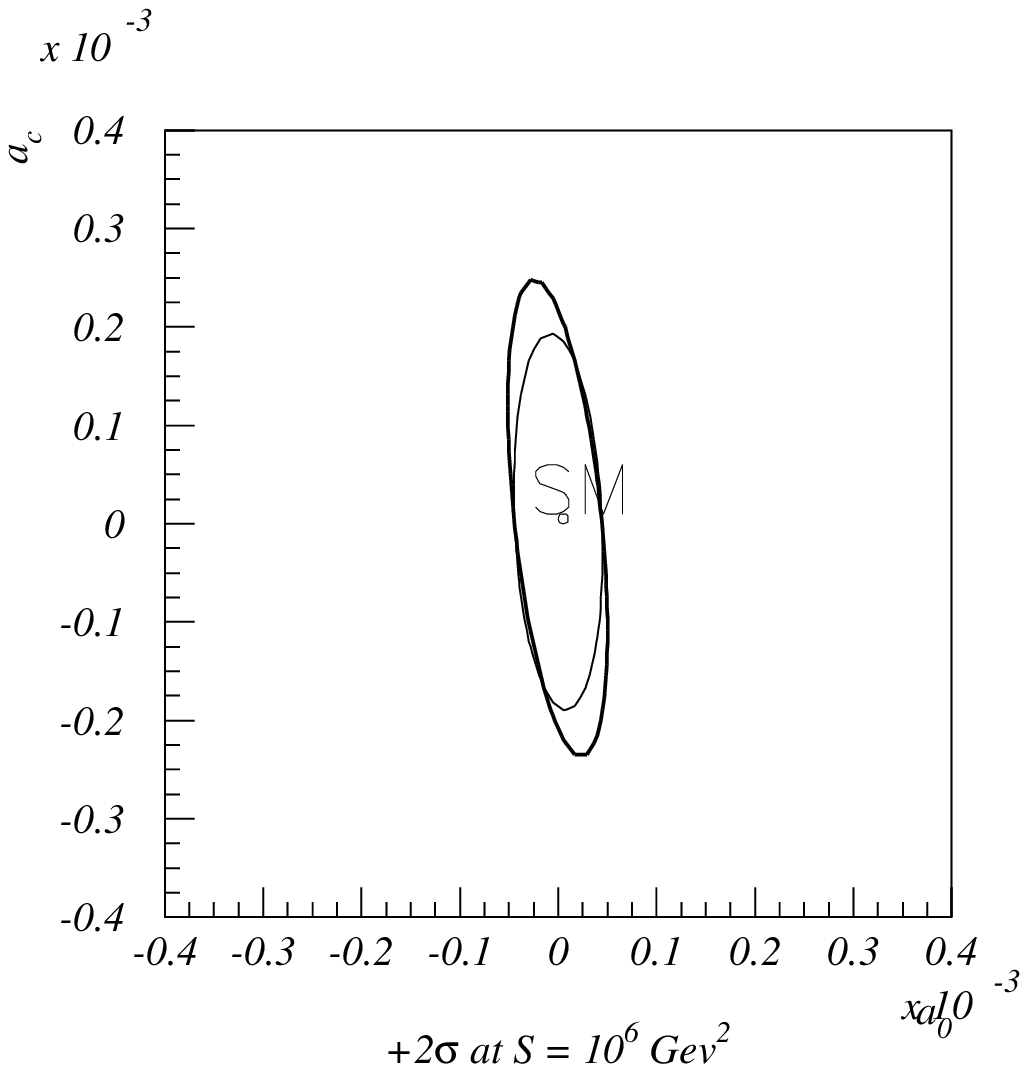}
    \hspace{0.1cm}
    \includegraphics[width=8.5cm]{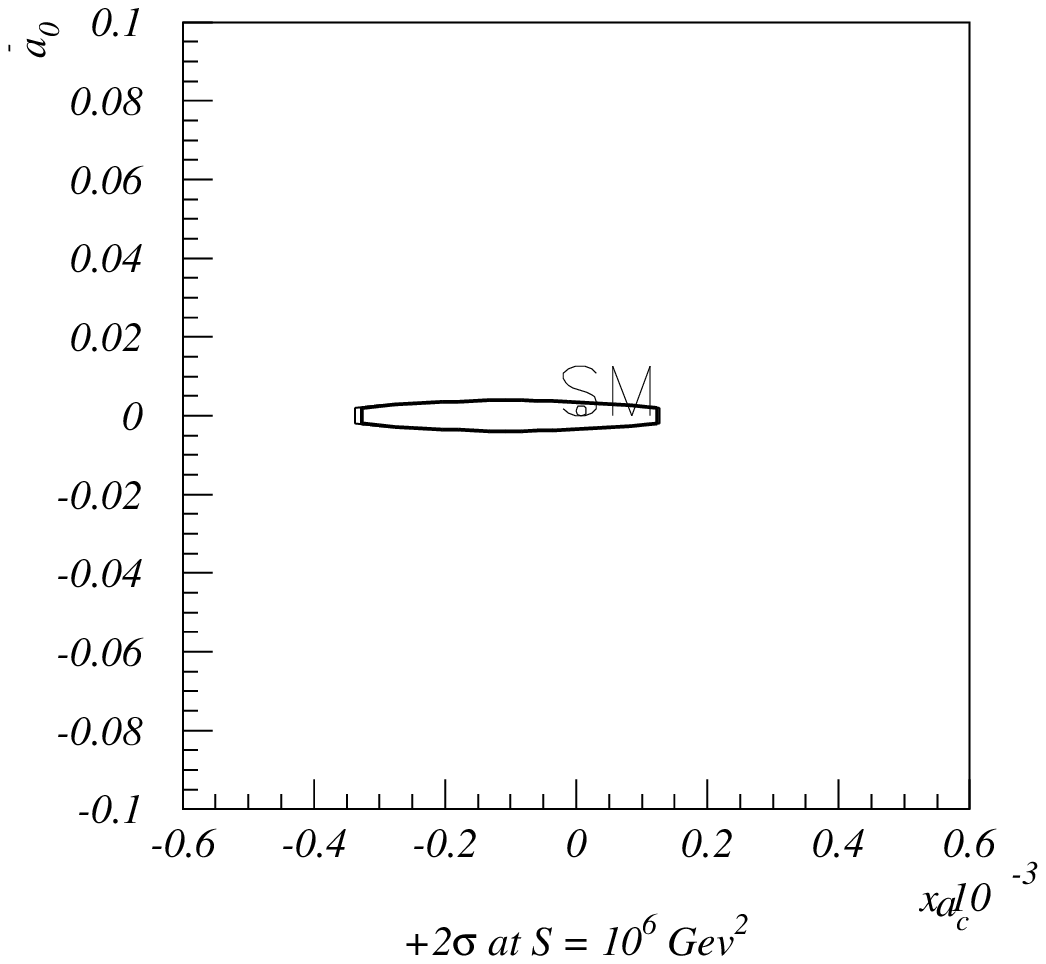}
    }
  }
 \caption{\it
      Contour plots for +2$\sigma$  deviation of $\sigma(\gamma\gamma\rightarrow W^+W^-)$: 
 with correction (single line) and without correction (double line)
    \label{p2} }
\end{figure}

Explicit expressions for the amplitude of the process $\gamma\gamma\rightarrow W^+W^-$ and $\gamma\gamma\rightarrow W^+W^-Z$ 
are contained in \cite{c9}. We suppose the following experimental conditions: unpolarized $\gamma\gamma$ beams at
$\sqrt{s}=1$ TeV with luminosity ${\cal L}=100fb^{-1}/year$. The contour plots for different deviations from the SM
$\gamma\gamma\rightarrow W^+W^-$ total cross sections at $\sqrt{s} = 1$ TeV, in case two of the three
anomalous couplings, are the most vivid and interesting for investigation of anomalous couplings. 
Figures \ref{p1} presents contour plots for different pairs of $a_i$. Here $\sigma$ means a variance of total cross section.
Figures \ref{p2} show the same contour plots with  radiative correction. 
The first order radiative correction  improves conditions of experiments 
concerning  anomalous constant. 

\end{document}